\documentclass [11pt,a4paper] {article}

\usepackage[cp1252]{inputenc}
\usepackage{amssymb}
\usepackage{amsmath}
\usepackage{amsfonts,amssymb}
\usepackage[dvips]{graphicx}
\usepackage{bbm}
\usepackage{enumerate}
\usepackage{amsthm}
\usepackage{cancel}

\DeclareMathAlphabet{\mathpzc}{OT1}{pzc}{m}{it}

\def\SmallColSep{\setlength{\arraycolsep}{1pt}}

\begin{document}

\title{Algebraic assignments of truth values to experimental quantum propositions}

\author{Arkady Bolotin\footnote{$Email: arkadyv@bgu.ac.il$\vspace{5pt}} \\ \emph{Ben-Gurion University of the Negev, Beersheba (Israel)}}

\maketitle

\begin{abstract}\noindent Of what are experimental quantum propositions primary bearers? As it is widely accepted in the modern literature, rather than being bearers of truth and falsity, these entities are bearers of probability values. Consequently, their truth values can be regarded as no more than degenerate probabilities (i.e., ones that have only the values 0 and 1). The mathematical motivation for precedence of probabilistic semantics over propositional semantic for the logic of experimental quantum propositions is Gleason’s theorem. It proves that the theory of probability measures on closed linear subspaces of a Hilbert space (which represent experimental quantum propositions) does not admit any probability measure having only the values 0 and 1.\\[-8pt]

\noindent By contrast, in the present paper, it is proclaimed that experimental propositions about quantum systems are primary bearers of truth values. As this paper demonstrates, algebraic properties of separable Hilbert spaces of finite dimension equal or greater than 2 do not allow in valuations (that is, truth assignments) which are dispersion-free, i.e., total functions from the set of atomic propositions to the set of two objects, true and false. Providing a probability function can be interpreted as a measure of the (un)certainty in the assignment of truth values, the fact that valuations cannot be dispersion-free gives rise to probabilistic semantics for the logic of experimental quantum propositions.\\

\end{abstract}

\section{Introduction}  %{<-------------------------------------------------------------------------------------------------Section I}

\noindent The relations between logic, probability theory and quantum mechanics are still a matter for investigation. One of the open questions is this: Among \emph{truth} and \emph{probability} pertaining to quantum systems, what is prior to what?\\

\noindent In the literature (see \cite{Wilce, Beltrametti, Chiara, Sorkin} to name but a few), it has been mainly accepted that \emph{quantum probability precedes quantum truth}. That is, rather than being bearers of truth values, experimental quantum propositions – i.e., meaningful declarative sentences that are (or make) statements about a quantum system – are primary bearers of probability values. This means that the logic of the experimental quantum propositions presupposes the probability theory of these entities. In other words, probabilistic semantics underlies propositional semantic for the logic of the experimental quantum propositions. As a result, valuations $v$ denoted by\smallskip

\begin{equation}  %{Eq.1}
   v
   \mkern-3.3mu
   :
   \mkern2mu
   \mathbb{P}
   \to
   \mathbb{B}_{2}
   \;\;\;\;  ,
\end{equation}
\smallskip

\noindent where $\mathbb{P}$ is the set of atomic propositions and $\mathbb{B}_{2}$ denotes either the set $\{\text{true} ,\text{false}\}$ or the set $\{1,0\}$, are replaced with probability functions $\Pr$ mapping elements of $\mathbb{P}$ to real numbers in the interval $[0,1]$ \smallskip

\begin{equation} \label{PR} %{Eq.2}
   \Pr
   \mkern-3.3mu
   :
   \mkern2mu
   \mathbb{P}
   \to
   \mathbb{R}
   \;\;\;\;  .
\end{equation}
\smallskip

\noindent Unlike axiomatic probability theory, in which probability functions are usually defined on $\sigma$-algebra of subsets of a given sample space $\Omega$, in the formula (\ref{PR}), functions $\Pr$ are defined on elements of the set $\mathbb{P}$ (the exposition of the notion of a probability function for the propositional language, which has a countable set of atomic propositions and the usual truth-functional connectives, see in \cite{Hajek, Williamson}). Therefore, the image of every proposition $P\in\mathbb{P}$ under any function $\Pr$, denoted $\Pr(P)$, is the probability of the proposition $P$ having the truth value of 1 in the given pure state of the system. Along these lines, valuations $v$ are merely degenerate probability functions Pr, namely,\smallskip

\begin{equation}  %{Eq.3}
   {[\mkern-3.3mu[P]\mkern-3.3mu]}_v
   =
   \left\{
      \begingroup\SmallColSep
      \begin{array}{r l}
         1
         ,
         &
         \mkern15mu
         \Pr(P)
         =
         1
         \\
         0
         ,
         &
         \mkern15mu
         \Pr(P)
         =
         0
      \end{array}
      \endgroup   
   \right.
   \;\;\;\;  ,
\end{equation}
\smallskip

\noindent where the double-bracket notation $ {[\mkern-3.3mu[\cdot]\mkern-3.3mu]}_v$ is used to express $v(\cdot)$.\\

\noindent Since, in accordance with Birkhoff and von Neumann’s proposal \cite{Birkhoff}, the mathematical representative of any experimental proposition about the quantum system is a closed linear subspace of a Hilbert space $\mathcal{H}$ associated with the system, it follows that probability functions $\Pr$ are defined on closed linear subspaces of $\mathcal{H}$. Providing the space $\mathcal{H}$ is separable (meaning that it admits an orthonormal basis consisting of a denumerable family of vectors), one can prove that $\Pr$ are continuous functions. Because no continuous function in the interval $[0,1]$ can take only the two values 0 and 1, the continuity of $\Pr$ implies that experimental quantum propositions cannot admit any $\Pr$ having only the values 0 and 1. This constitutes Gleason's theorem \cite{Gleason}.\\

\noindent However, the question of precedence between quantum truth and quantum probability might have a different answer.\\

\noindent Indeed, imagine that $v$, a function from $\mathbb{P}$ to $\mathbb{B}_{2}$, is not total but \emph{partial}. In that case, for any proposition $P\in\mathbb{P}$, either $ {[\mkern-3.3mu[P]\mkern-3.3mu]}_v$ belongs to $\mathbb{B}_{2}$ or $ {[\mkern-3.3mu[P]\mkern-3.3mu]}_v$ is undefined (i.e., the proposition $P$ has something called \emph{truth-value gap} \cite{Beziau}, meaning that $P$ is neither true nor false). In addition, assume that there exists a probability function $\Pr$ from $\mathbb{P}$ to $\mathbb{R}$ mapping each proposition $P$ to $\Pr(P)\in[0,1]$ in a way that ${[\mkern-3.3mu[P]\mkern-3.3mu]}_v \neq {[\mkern-3.3mu[P^\prime]\mkern-3.3mu]}_v$ implies $\Pr(P)\neq\Pr(P^\prime)$, and also\smallskip

\begin{equation}  %{Eq.4}
   \Pr(P)
   =
   \left\{
      \begingroup\SmallColSep
      \begin{array}{r l}
         1
         ,
         &
         \mkern15mu
         {[\mkern-3.3mu[P]\mkern-3.3mu]}_v
         =
         1
         \\
         0
         ,
         &
         \mkern15mu
         {[\mkern-3.3mu[P]\mkern-3.3mu]}_v
         =
         0
      \end{array}
      \endgroup   
   \right.
   \;\;\;\;  .
\end{equation}
\smallskip

\noindent Then, in case ${[\mkern-3.3mu[P]\mkern-3.3mu]}_v$ is neither 1 nor 0, it must be that $\Pr(P)\notin\{0,1\}$.\\

\noindent Now, consider the following mathematical statement (which can be called the lemma of partial valuation): Without any additional supposition, algebraic properties of separable Hilbert spaces $\mathcal{H}$ cannot allow in truth assignments  $v$ which are total functions from $\mathbb{P}$ to $\mathbb{B}_{2}$. If this statement is true, it will enable the same inference as Gleason’s theorem does – that experimental quantum propositions do not admit probabilities $\Pr(P)$ which are only $\{0,1\}$-valued.\\

\noindent The present paper provides the proof of the above statement.\\

\section{Truth assignments in agreement with quantum theory}  %{<-------------------------------------------------------------------------------------------------Section II}

\noindent Recall that any closed linear subspace of $\mathcal{H}$ is \emph{the range} of some projection operator $\hat{P}$ acting on $\mathcal{H}$ \cite{Kalmbach}, explicitly,\smallskip

\begin{equation}  %{Eq.5}
   \mathrm{ran}(\hat{P})
   =
   \left\{
      |\psi\rangle
      \in
      \mathcal{H}
      \textnormal{:}
      \mkern10mu
      \hat{P}
      |\psi\rangle
      =
      |\psi\rangle      
   \right\}
   \;\;\;\;  .
\end{equation}
\smallskip

\noindent In view of that, the zero subspace $\{0\}$ and the identity subspace $\mathcal{H}$ are \emph{trivial closed linear subspaces} of $\mathcal{H}$, namely, $\{0\}=\mathrm{ran}(\hat{0})$ and $\mathcal{H}=\mathrm{ran}(\hat{1})$, where $\hat{0}$ and $\hat{1}$ are the zero and identity operators, correspondingly.\\

\noindent Because the set of the eigenvalues of each projection operator $\hat{P}$ is contained in $\{0,1\}$, one can assume correspondence between an experimental proposition $P$ and a projection operator $\hat{P}$, which is another way of stating that the mathematical representative of an experimental proposition $P$ is a nontrivial closed linear subspace $\mathrm{ran}(\hat{P})$.\\

\noindent Let the system be in the pure state described by the unit vector $|\Psi\rangle\in\mathcal{H}$, i.e., one that has unit norm, $\langle\Psi|\Psi\rangle=1$. Then, the truth value of the experimental proposition $P$ in the state $|\Psi\rangle$ can be determined by the formula\smallskip

\begin{equation}  %{Eq.6}
   {[\mkern-3.3mu[
      P
   ]\mkern-3.3mu]}_v
   =
   \mathfrak{P}_{\in}
   \!
   \left(
      |\Psi\rangle
      ,
      \mathrm{ran}(\hat{P})
   \right)
   \;\;\;\;  ,
\end{equation}
\smallskip

\noindent where $\mathfrak{P}_{\in}(|\Psi\rangle, \mathrm{ran}(\hat{P}))$ is the image of a couple $(|\Psi\rangle, \mathrm{ran}(\hat{P}))$ under the propositional function (in other words, \emph{predicate}) $\mathfrak{P}_{\in}$ denoted by the mapping\smallskip

\begin{equation}  %{Eq.7}
   \mathfrak{P}_{\in}
   :
   \mathfrak{H}
   \times
   \wp({\mathcal{H}})
   \to
   \mathbb{B}_{2}
   \;\;\;\;   ,
\end{equation}
\smallskip

\noindent in which $\mathfrak{H}$ is the set of all unit vectors in $\mathcal{H}$ and $\wp({\mathcal{H}})$ is the set of all closed linear subspaces of $\mathcal{H}$. The predicate $\mathfrak{P}_{\in}$ is used to indicate set membership: $\mathfrak{P}_{\in}(|\Psi\rangle,\mathrm{ran}(\hat{P}))=1$ if the vector $|\Psi\rangle$ belongs to $\mathrm{ran}(\hat{P})$; contrastively, $\mathfrak{P}_{\in}(|\Psi\rangle,\mathrm{ran}(\hat{P}))=0$ if $|\Psi\rangle$ does not belong to $\mathrm{ran}(\hat{P})$.\\

\noindent Consider \emph{the kernel} of the projection operator $\hat{P}$: It is the closed linear subspace of $\mathcal{H}$ that corresponds the set of vectors $|\phi\rangle$ in $\mathcal{H}$ which are mapped to zero by $\hat{P}$, i.e.,\smallskip

\begin{equation}  %{Eq.8}
   \mathrm{ker}(\hat{P})
   =
   \mathrm{ran}(\hat{1}-\hat{P})
   =
   \left\{
      |\phi\rangle
      \in
      \mathcal{H}
      \textnormal{:}
      \mkern10mu
      \hat{P}
      |\phi\rangle
      =
      0      
   \right\}
   \;\;\;\;  .
\end{equation}
\smallskip

\noindent As every unit vector $|\Psi\rangle\in\mathcal{H}$ can be decomposed uniquely as $|\Psi\rangle=|\psi\rangle+|\phi\rangle$ with $|\psi\rangle=\hat{P}|\Psi\rangle$ and $|\phi\rangle=|\Psi\rangle-\hat{P}|\Psi\rangle=(\hat{1}-\hat{P})|\Psi\rangle$, where $|\psi\rangle\in\mathrm{ran}(\hat{P})$ and $|\phi\rangle\in\mathrm{ker}(\hat{P})$, the subspaces $\mathrm{ran}(\hat{P})$ and $\mathrm{ker}(\hat{P})$ decompose the Hilbert space $\mathcal{H}$ into the direct sum:\smallskip

\begin{equation}  %{Eq.9}
   \mathcal{H}
   =
   \mathrm{ran}(\hat{P})
   \oplus
   \mathrm{ker}(\hat{P})
   =
   \mathrm{ran}(\hat{P})
   \oplus
   \mathrm{ran}(\hat{1}-\hat{P})
   \;\;\;\;  .
\end{equation}
\smallskip

\noindent Since $|\Psi\rangle\notin\{0\}$, one can infer from here that\smallskip

\begin{equation}  %{Eq.10}
   |\Psi\rangle
   \in
   \mathrm{ker}
      (\hat{P})
   \implies
   |\Psi\rangle
   \notin
   \mathrm{ran}
      (\hat{P})
   \;\;\;\;  ,
\end{equation}
\smallskip

\noindent i.e., if $|\Psi\rangle$ belongs to $\mathrm{ker}(\hat{P})$, then $|\Psi\rangle$ does not belong to $\mathrm{ran}(\hat{P})$ and so $\mathfrak{P}_{\in}(|\Psi\rangle,\mathrm{ran}(\hat{P}))=0$.\\ 

\noindent But suppose $|\Psi\rangle$ belongs to neither $\mathrm{ran}(\hat{P})$ nor $\mathrm{ker}(\hat{P})$, i.e., both $|\Psi\rangle\mkern-2mu\notin\mkern-2mu\mathrm{ran}(\hat{P})$ and $|\Psi\rangle\mkern-2mu\notin\mkern-2mu\mathrm{ker}(\hat{P})$ are true. This is logically equivalent to truth of \emph{the joint denial} $|\Psi\rangle\mkern-2mu\in\mkern-2mu\mathrm{ran}(\hat{P})\mkern-2mu\downarrow\mkern-2mu|\Psi\rangle\mkern-2mu\in\mkern-2mu\mathrm{ker}(\hat{P})$ (recall that the joint denial or \emph{logical nor} is a truth-functional operator which produces a result that is the negation of \emph{logical or}, $\sqcup$).\\

\noindent To take into consideration this case, one can put forward the hypothesis of indistinguishability:\\

\noindent \textbf{Hypothesis:} \emph{The statements $|\Psi\rangle\mkern-2mu\in\mkern-2mu\mathrm{ker}(\hat{P})$ and $|\Psi\rangle\mkern-2mu\in\mkern-2mu\mathrm{ran}(\hat{P})\mkern-2mu\downarrow\mkern-2mu|\Psi\rangle\mkern-2mu\in\mkern-2mu\mathrm{ker}(\hat{P})$ are indistinguishable from one another under the propositional function $\mathfrak{P}_{\in}$}.\\

\noindent In a semantics defined upon this hypothesis, $\mathfrak{P}_{\in}$ is a total function and presented by\smallskip

\begin{equation} \label{QLOG} %{Eq.11}
   \mathfrak{P}_{\in}
   \!
   \left(
      |\Psi\rangle
      ,
      \mathrm{ran}(\hat{P})
   \right)
   =
   \left\{
      \begingroup\SmallColSep
      \begin{array}{r l}
         1
         ,
         &
         \mkern15mu
         |\Psi\rangle
         \in
         \mathrm{ran}
            (\hat{P})
         \\
         0
         ,
         &
         \mkern15mu
         |\Psi\rangle
         \in
         \mathrm{ker}
            (\hat{P})
         \\
         0
         ,
         &
         \mkern15mu
         |\Psi\rangle
         \in
         \mathrm{ran}
            (\hat{P})
         \downarrow
         |\Psi\rangle
         \in
         \mathrm{ker}
            (\hat{P})
      \end{array}
      \endgroup   
   \right.
   \;\;\;\;  .
\end{equation}
\smallskip

\noindent Despite being bivalent, such a semantics is not classical since it does not hold the distributive law of classical logic. Let us show this.\\

\noindent Assume that $\neg{P}$, the negation of the proposition $P$, corresponds to the projection operator $\hat{1}-\hat{P}$, and the disjunction $P\sqcup\neg{P}$ is represented by the lattice-theoretic join $\mathrm{ran}(\hat{P})\vee\mathrm{ran}(\hat{1}-\hat{P})$. This gives\smallskip

\begin{equation}  %{Eq.12}
   \mathrm{ran}(\hat{P})
   \vee
   \mathrm{ran}(\hat{1}-\hat{P})
   =
   \mathrm{ran}(\hat{P})
   \oplus
   \mathrm{ker}(\hat{P})
   =
   \mathcal{H}
   \;\;\;\;  .
\end{equation}
\smallskip

\noindent Let the unit vector $|\Phi\rangle\mkern-3.5mu\in\mkern-3.5mu\mathcal{H}$ be such that it belongs to the subspace $\mathrm{ran}(\hat{Q})$, which represents the proposition $Q$, but with this, $|\Phi\rangle$ does not belong to neither $\mathrm{ran}(\hat{P})$ nor $\mathrm{ran}(\hat{1}-\hat{P})$. Then, according to (\ref{QLOG}),\smallskip

\begin{equation}  %{Eq.13}
   \mathfrak{P}_{\in}(|\Phi\rangle,\mathrm{ran}(\hat{Q}))
   =
   1
   \;\;\;\;  ,
\end{equation}
\\[-35pt]

\begin{equation}  %{Eq.14}
   \mathfrak{P}_{\in}(|\Phi\rangle,\mathrm{ran}(\hat{P}))
   =
   0
   \;\;\;\;  ,
\end{equation}
\\[-35pt]

\begin{equation}  %{Eq.15}
   \mathfrak{P}_{\in}(|\Psi\rangle,\mathrm{ker}(\hat{P}))
   =
   0
   \;\;\;\;  .
\end{equation}
\smallskip

\noindent Let us also assume that the lattice-theoretic meet $\mathrm{ran}(\hat{Q})\wedge\mathrm{ran}(\hat{P})$ is the set-theoretic intersection $\mathrm{ran}(\hat{Q})\cap\mathrm{ran}(\hat{P})$. In the set builder notation, the latter is written down as:\smallskip

\begin{equation}  %{Eq.16}
   \mathrm{ran}(\hat{Q})
   \cap
   \mathrm{ran}(\hat{P})
   =
   \Big\{
      |\psi\rangle
      \in
      \mathcal{H}
      \textnormal{:}
      \mkern10mu
      \mathfrak{F}
      \left(
         |\psi\rangle
      \right)
      \mkern-3mu
   \Big\}
   \;\;\;\;  ,
\end{equation}
\smallskip

\noindent where the rule $\mathfrak{F}(\psi\rangle)$ is the logical conjunction of two predicates, namely,\smallskip

\begin{equation}  %{Eq.17}
   \mathfrak{F}
   \left(
      |\psi\rangle
   \right)
   =
   \mathfrak{P}_{\in}
   \!
   \left(
      |\psi\rangle
      ,
      \mathrm{ran}(\hat{Q})
   \right)
   \mkern-2mu
   \sqcap
   \mathfrak{P}_{\in}
   \!
   \left(
      |\psi\rangle
      ,
      \mathrm{ran}(\hat{P})
   \right)
   \;\;\;\;  .
\end{equation}
\smallskip

\noindent If $|\psi\rangle=|\Phi\rangle$, then $\mathfrak{F}(|\Phi\rangle)=0$, and so the non-zero vector $|\Phi\rangle$ is not an element of $\mathrm{ran}(\hat{Q})\wedge\mathrm{ran}(\hat{P})$. This indicates that $\mathrm{ran}(\hat{Q})\wedge\mathrm{ran}(\hat{P})=\{0\}$ and, likewise, $\mathrm{ran}(\hat{Q})\wedge\mathrm{ran}(\hat{1}-\hat{P})=\{0\}$.\\

\noindent Providing $(Q\sqcap{P})\sqcup(Q\sqcap\neg{P})$ is represented by $(\mathrm{ran}(\hat{Q})\mkern-1mu\wedge\mathrm{ran}(\hat{P}))\vee(\mathrm{ran}(\hat{Q})\mkern-1mu\wedge\mathrm{ran}(\hat{1}-\hat{P}))=\{0\}\vee\{0\}=\{0\}$, and $Q\sqcap(P\sqcup\neg{P})$ is represented by $\mathrm{ran}(\hat{Q})\wedge\mathcal{H}=\mathrm{ran}(\hat{Q})$, the failure of the distributive law $Q\sqcap(P\sqcup\neg{P}) = (Q\sqcap{P})\sqcup(Q\sqcap\neg{P})$ ensues:\smallskip

\begin{equation}  %{Eq.18}
   {[\mkern-3.3mu[
      \mkern1.5mu
      Q\sqcap(P\sqcup\neg{P})
   ]\mkern-3.3mu]}_v
   =
   \mathfrak{P}_{\in}(|\Phi\rangle,\mathrm{ran}(\hat{Q}))
   =
   1
   \;\;\;\;  ,
\end{equation}
\smallskip

\noindent however\smallskip

\begin{equation}  %{Eq.19}
   {[\mkern-3.3mu[
      (Q\sqcap{P})\sqcup(Q\sqcap\neg{P})
   ]\mkern-3.3mu]}_v
   =
   \mathfrak{P}_{\in}(|\Phi\rangle,\{0\})
   =
   0
   \;\;\;\;  .
\end{equation}
\smallskip

\noindent The described semantics is identified with quantum logic (of Birkhoff and von Neumann \cite{Birkhoff}).\\

\noindent It is important to lay stress on the fact that the hypothesis of indistinguishability stated above is neither intuitive nor plausible nor justifiable by experimental evidence. This hypothesis appears to be added just for the purpose of maintaining valuations $v$ as total functions from $\mathbb{P}$ to $\mathbb{B}_2$.\\

\noindent Hence, \emph{if a semantics is presumed bivalent but no further hypothesis is assumed}, the predicate $\mathfrak{P}_{\in}$ can only be a partial function, that is,\smallskip

\begin{equation} \label{SV} %{Eq.20}
   \mathfrak{P}_{\in}
   \!
   \left(
      |\Psi\rangle
      ,
      \mathrm{ran}(\hat{P})
   \right)
   =
   \left\{
      \begingroup\SmallColSep
      \begin{array}{r l}
         1
         ,
         &
         \mkern15mu
         |\Psi\rangle
         \in
         \mathrm{ran}
            (\hat{P})
         \\
         0
         ,
         &
         \mkern15mu
         |\Psi\rangle
         \in
         \mathrm{ker}
            (\hat{P})
         \\
         0/0
         ,
         &
         \mkern15mu
         |\Psi\rangle
         \in
         \mathrm{ran}
            (\hat{P})
         \downarrow
         |\Psi\rangle
         \in
         \mathrm{ker}
            (\hat{P})
      \end{array}
      \endgroup   
   \right.
   \;\;\;\;  ,
\end{equation}
\smallskip

\noindent where $0/0$ symbolizes a truth-value gap.\\ 

\noindent In a semantics of this kind, $\mathfrak{P}_{\in}(|\Phi\rangle,\mathrm{ran}(\hat{P}))=0/0$; therefore, the rule $\mathfrak{F}(|\Phi\rangle)$ cannot be determined, which causes the meet of two subspaces $\mathrm{ran}(\hat{Q})$ and $\mathrm{ran}(\hat{P})$ to be \emph{undecidable}.\\

\noindent What is more, in this semantics, the disjunction $P\sqcup\neg{P}$ is true even in the case where neither $P$ nor $\neg{P}$ has a truth value: To be sure, ${[\mkern-3.3mu[P\sqcup\neg{P})]\mkern-3.3mu]}_v = \mathfrak{P}_{\in}(|\Phi\rangle,\mathcal{H}) = 1$ even though ${[\mkern-3.3mu[P]\mkern-3.3mu]}_v = \mathfrak{P}_{\in}(|\Phi\rangle,\mathrm{ran}(\hat{P})) = 0/0$ and ${[\mkern-3.3mu[\neg{P}]\mkern-3.3mu]}_v = \mathfrak{P}_{\in}(|\Phi\rangle,\mathrm{ran}(\hat{1}-\hat{P})) = 0/0$.\\

\noindent The above semantics is identified with \emph{supervaluationism}, i.e., the form of \emph{partial logic} (a deeper study of truth-value gaps and logics that allow for truth-value gaps can be found, for example, in \cite{Blamey, Langholm}).\\

\noindent As one can see from (\ref{SV}), the logic of experimental quantum propositions could have been classical (i.e., total and bivalent), if the joint denial $|\Psi\rangle\mkern-2mu\in\mkern-2mu\mathrm{ran}(\hat{P})\mkern-2mu\downarrow\mkern-2mu|\Psi\rangle\mkern-2mu\in\mkern-2mu\mathrm{ker}(\hat{P})$ would have been \emph{always false}, i.e., false for all couples $(|\Psi\rangle, \mathrm{ran}(\hat{P}))$.\\

\noindent Let us clarify the reason why this condition was not fulfilled.\\

\section{Set membership from the algebraic perspective}  %{<-------------------------------------------------------------------------------------------------Section III}

\noindent Consider a separable Hilbert space $\mathcal{H}$ of finite dimension $n$. Let the projection operator $\hat{P}$ acting on $\mathcal{H}=\mathbb{C}^{n}$ be expressed in terms of the complex ${n}\times{n}$ matrix $\mathbf{P}$\smallskip

\begin{equation}  %{Eq.21}
   \mathbf{P}
   =
   \!\left[
      \begingroup\SmallColSep
      \begin{array}{c c c}
         P_{11} & \cdots  & P_{1n} \\
         \vdots & \ddots & \vdots \\
         P_{n1} & \cdots  & P_{nn}
      \end{array}
      \endgroup
   \right]
   =
   \left(
      P_{ij}
   \right)_{i=1,\mkern2mu{j=1}}^{n,n}
   \in
   \mathbf{Mat_{n\times{n}}}(\mathbb{C})
   \;\;\;\;  .
\end{equation}
\smallskip

\noindent Then, the range of $\hat{P}$ is the same as the span of the column vectors $\mathbf{P}\mkern-3mu_{j}$ of the matrix $\mathbf{P}$, i.e.,\smallskip

\begin{equation}  %{Eq.22}
   \mathrm{ran}(\hat{P})
   =
   \mathrm{Span}
   \left(
      \mathbf{P}\mkern-3mu_{1}
      ,
      \dots
      ,
      \mathbf{P}\mkern-3mu_{n}
   \right)
   \;\;\;\;  ,
\end{equation}
\smallskip

\noindent where either $(\mathbf{P}\mkern-3mu_{1},\dots,\mathbf{P}\mkern-3mu_{n})$ is a basis of $\mathrm{ran}(\hat{P})$ or some $\mathbf{P}\mkern-3mu_{j}$ can be removed to obtain a basis of $\mathrm{ran}(\hat{P})$; explicitly,\smallskip

\begin{equation}  %{Eq.23}
   \mathrm{ran}(\hat{P})
   =
   \left\{
      c_1
      ,
      \dots
      c_n
      \in
      \mathbb{C}
      :
      \mkern15mu
      c_1
      \mkern-4mu
      \left[
         \begingroup\SmallColSep
         \begin{array}{c}
            P_{11}  \\
            \vdots  \\
            P_{n1} 
         \end{array}
         \endgroup
      \right]
      +
      \cdots
      +
      c_n
      \mkern-4mu
      \left[
         \begingroup\SmallColSep
         \begin{array}{c}
            P_{1n}  \\
            \vdots  \\
            P_{nn} 
         \end{array}
         \endgroup
      \right]
   \right\}
   \;\;\;\;  .
\end{equation}
\smallskip

\noindent Provided a basis set of vectors $\{|e_1\rangle,\dots,|e_n\rangle\}$ in the Hilbert space $\mathbb{C}^{n}$, the unit vector $|\Psi\rangle$ describing the pure state of the quantum system can be expressed as the column vector $\mathbf{\Psi}\in\mkern-2mu\mathbf{Mat_{n\times{1}}}(\mathbb{C})$ whose $i^{\text{th}}$ row has the entry $\langle{e_i}|\Psi\rangle$.\\

\noindent One can make here the following observation: From the algebraic perspective, the truth of the statement $|\Psi\rangle\mkern-2mu\in\mkern-2mu\mathrm{ran}(\hat{P})$ entails the existence of at least one solution to the system of linear equations\smallskip

\begin{equation}  %{Eq.24}
   \mathbf{RX}
   =
   \mathbf{\Psi}
    \;\;\;\;  ,
\end{equation}
\smallskip

\noindent where $\mathbf{X}\in\mkern-2mu\mathbf{Mat_{m\times{1}}}(\mathbb{C})$ is the column vector with $m\le{n}$ unknowns $x_1,\dots,x_m$ which are put in the place of weights $c_1,\dots,c_m$ for the linearly independent column vectors $\mathbf{P}\mkern-3mu_{1},\dots,\mathbf{P}\mkern-3mu_{m}$ of the matrix $\mathbf{P}$, so that\smallskip

\begin{equation}  %{Eq.25}
   \mathbf{RX}
   =
   \left[
      \begingroup\SmallColSep
      \begin{array}{c}
         P_{11}  \\
         \vdots  \\
         P_{n1} 
      \end{array}
      \endgroup
   \right]
   \mkern-4mu
   x_1
   +
   \cdots
   +
   \left[
      \begingroup\SmallColSep
      \begin{array}{c}
         P_{1m}  \\
         \vdots  \\
         P_{nm} 
      \end{array}
      \endgroup
   \right]
   \mkern-4mu
   x_m
    \;\;\;\;  .
\end{equation}
\smallskip

\noindent Denoting $(\delta_{ij})_{i=1}^{n}\in\mathbf{Mat_{n\times{1}}}(\mathbb{C})$ by $\mathbf{I}_{j}$, the kernel of $\hat{P}$ can be presented as the span of the column vectors $\mathbf{I}_j-\mathbf{P}_j$ of the matrix $\mathbf{I}-\mathbf{P}$, that is,\smallskip

\begin{equation}  %{Eq.26}
   \mathrm{ker}(\hat{P})
   =
   \mathrm{Span}
   \left(
      \mathbf{I}_{1}
      -
      \mathbf{P}\mkern-3mu_{1}
      ,
      \dots
      ,
      \mathbf{I}_{n}
      -
      \mathbf{P}\mkern-3mu_{n}
   \right)
   \;\;\;\;  ,
\end{equation}
\smallskip

\noindent where either $(\mathbf{I}_{1}-\mathbf{P}\mkern-3mu_{1},\dots,\mathbf{I}_{n}-\mathbf{P}\mkern-3mu_{n})$ is a basis of $\mathrm{ker}(\hat{P})$ or some $\mathbf{I}_{j}-\mathbf{P}\mkern-3mu_{j}$ can be removed to obtain a basis of $\mathrm{ker}(\hat{P})$; explicitly,\smallskip

\begin{equation}  %{Eq.27}
   \mathrm{ker}(\hat{P})
   =
   \left\{
      c_1
      ,
      \dots
      c_n
      \in
      \mathbb{C}
      :
      \mkern15mu
      c_1
      \mkern-4mu
      \left[
         \begingroup\SmallColSep
         \begin{array}{r}
            1-P_{11}  \\
            \vdots     \\
            -P_{n1} 
         \end{array}
         \endgroup
      \right]
      +
      \cdots
      +
      c_j
      \mkern-4mu
      \left[
         \begingroup\SmallColSep
         \begin{array}{r}
            \vdots                 \\
            \delta_{ij}-P_{ij}  \\
            \vdots 
         \end{array}
         \endgroup
      \right]
      +
      \cdots
      +
      c_n
      \mkern-4mu
      \left[
         \begingroup\SmallColSep
         \begin{array}{r}
            -P_{1n}  \\
            \vdots   \\
            1-P_{nn} 
         \end{array}
         \endgroup
      \right]
   \right\}
   \;\;\;\;  .
\end{equation}
\smallskip

\noindent Accordingly, to decide whether $|\Psi\rangle$ belongs to $\mathrm{ker}(\hat{P})$ means to answer the question whether the following system of linear equations has at least one solution:\smallskip

\begin{equation}  %{Eq.28}
   \mathbf{KX}
   =
   \mathbf{\Psi}
    \;\;\;\;  ,
\end{equation}
\smallskip

\noindent where $\mathbf{X}\in\mkern-2mu\mathbf{Mat_{k\times{1}}}(\mathbb{C})$ is the column vector with $k\le{n}$ unknowns $x_1,\dots,x_k$ which substitute weights $c_1,\dots,c_k$ for the linearly independent column vectors of the matrix $\mathbf{I}-\mathbf{P}$ so that\smallskip

\begin{equation}  %{Eq.29}
   \mathbf{KX}
   =
   \left[
      \begingroup\SmallColSep
      \begin{array}{r}
         1-P_{11}  \\
         \vdots     \\
         -P_{n1} 
      \end{array}
      \endgroup
   \right]
   \mkern-4mu
   x_1
   +
   \cdots
   +
   \left[
      \begingroup\SmallColSep
      \begin{array}{r}
         \vdots                 \\
         \delta_{ij}-P_{ij}  \\
         \vdots 
      \end{array}
      \endgroup
   \right]
   \mkern-4mu
   x_j
   +
   \cdots
   +
   \left[
      \begingroup\SmallColSep
      \begin{array}{r}
         -P_{1n}  \\
         \vdots   \\
         1-P_{nk} 
      \end{array}
      \endgroup
   \right]
   \mkern-4mu
   x_k
    \;\;\;\;  .
\end{equation}
\smallskip

\noindent To illustrate this observation, consider the Hilbert space $\mathbb{C}^4$ characterizing the spin $^{3}\mkern-5mu/\mkern-3mu_{2}$ system. In terms of the complex $4\times4$ matrix, the projection operator $\hat{Y}_{+^{3}\mkern-5mu/\mkern-3mu_{2}}$ corresponding to the experimental atomic proposition “The spin of the system along the $Y$ axis is $+^{3}\mkern-5mu/\mkern-3mu_{2}\hbar\mkern3mu$”, denoted $Y_{+^{3}\mkern-5mu/\mkern-3mu_{2}}$, takes the form\smallskip

\begin{equation}  %{Eq.30}
   \mathbf{Y}_{+^{3}\mkern-5mu/\mkern-3mu_{2}}
   =
   \frac{1}{8}
   \left[
      \begingroup
      \begin{array}{r r r r}
                      1 & -i\sqrt{3}  & -\sqrt{3} &                i \\
         i\sqrt{3} &                3 &           -i3  &  -\sqrt{3} \\
         -\sqrt{3} &              i3  &              3 & -i\sqrt{3}  \\
                      -i &   -\sqrt{3} & i\sqrt{3} &               1
      \end{array}
      \endgroup
   \right]
   \;\;\;\;  ;
\end{equation}
\smallskip

\noindent its range and kernel are\smallskip

\begin{equation}  %{Eq.31}
   \mathrm{ran}(\hat{Y}_{+^{3}\mkern-5mu/\mkern-3mu_{2}})
   =
   \left\{
      a
      \in
      \mathbb{C}
      :
      \mkern15mu
      a
      \mkern-4mu
      \left[
         \begingroup\SmallColSep
         \begin{array}{r}
                          1 \\
            i\sqrt{3}  \\
            -\sqrt{3}  \\
                         -i
         \end{array}
         \endgroup
      \right]
   \right\}
   \;\;\;\;  ,
\end{equation}
\\[-25pt]

\begin{equation}  %{Eq.32}
   \mathrm{ker}(\hat{Y}_{+^{3}\mkern-5mu/\mkern-3mu_{2}})
   =
   \left\{
      a
      ,
      b
      ,
      c
      \in
      \mathbb{C}
      :
      \mkern15mu
      a
      \mkern-4mu
      \left[
         \begingroup\SmallColSep
         \begin{array}{r}
                           7 \\
            -i\sqrt{3}  \\
              \sqrt{3}  \\
                           i 
         \end{array}
         \endgroup
      \right]
      +
      b
      \mkern-4mu
      \left[
         \begingroup\SmallColSep
         \begin{array}{r}
             i\sqrt{3} \\
                          5 \\
                        -i3 \\
              \sqrt{3}
         \end{array}
         \endgroup
      \right]
      +
      c
      \mkern-4mu
      \left[
         \begingroup\SmallColSep
         \begin{array}{r}
                \sqrt{3} \\
                         i3 \\
                          5 \\
            -i\sqrt{3}
         \end{array}
         \endgroup
      \right]
   \right\}
   \;\;\;\;  .
\end{equation}
\smallskip

\noindent Suppose that the spin $^{3}\mkern-5mu/\mkern-3mu_{2}$ system is in the pure state that is described by the ket $|{Y}_{+^{3}\mkern-5mu/\mkern-3mu_{2}}\rangle$ expressed as the column vector written out in the coordinates over the $z$-basis, namely,\smallskip

\begin{equation}  %{Eq.33}
   |{Y}_{+^{3}\mkern-5mu/\mkern-3mu_{2}}\rangle
   =
   \frac{1}{2\sqrt{2}}
   \mkern-4mu
   \left[
      \begingroup\SmallColSep
      \begin{array}{r}
                     i  \\
          -\sqrt{3} \\
         -i\sqrt{3} \\
                    1
      \end{array}
      \endgroup
   \right]
   \;\;\;\;  .
\end{equation}
\smallskip

\noindent To decide whether the statement $|{Y}_{+^{3}\mkern-5mu/\mkern-3mu_{2}}\rangle\mkern-2mu\in\mkern-2mu\mathrm{ran}(\hat{Y}_{+^{3}\mkern-5mu/\mkern-3mu_{2}})$ is true, let us present it as the system of linear equations\smallskip

\begin{equation} \label{SYS1} %{Eq.34}
   \left[
      \begingroup\SmallColSep
      \begin{array}{r}
                       1 \\
         i\sqrt{3}  \\
         -\sqrt{3}  \\
                      -i
      \end{array}
      \endgroup
   \right]
   \mkern-4mu
   x
   =
   \frac{1}{2\sqrt{2}}
   \mkern-4mu
   \left[
      \begingroup\SmallColSep
      \begin{array}{r}
                        i  \\
          -\sqrt{3}  \\
         -i\sqrt{3}  \\
                       1
      \end{array}
      \endgroup
   \right]
   \;\;\;\;  .
\end{equation}
\smallskip

\noindent Even though this linear system is overdetermined, it contains 3 linearly dependent equations; hence, it has the solution, $x=i \frac{i\sqrt{2}}{4}$, which means that the ket $|{Y}_{+^{3}\mkern-5mu/\mkern-3mu_{2}}\rangle$ belongs to $\mathrm{ran}(\hat{Y}_{+^{3}\mkern-5mu/\mkern-3mu_{2}})$, and so in the state described by $|{Y}_{+^{3}\mkern-5mu/\mkern-3mu_{2}}\rangle$, the proposition ${Y}_{+^{3}\mkern-5mu/\mkern-3mu_{2}}$ is true:\smallskip

\begin{equation}  %{Eq.35}
   {[\mkern-3.3mu[
      {Y}_{+^{3}\mkern-5mu/\mkern-3mu_{2}}
   ]\mkern-3.3mu]}_v
   =
   \mathfrak{P}_{\in}
   \!
   \left(
      |{Y}_{+^{3}\mkern-5mu/\mkern-3mu_{2}}\rangle
      ,
      \mathrm{ran}(\hat{Y}_{+^{3}\mkern-5mu/\mkern-3mu_{2}})
   \right)
   =
   1
   \;\;\;\;  .
\end{equation}
\smallskip

\noindent Yet, the linear system with the same left-hand side as (\ref{SYS1}) but the different right-hand side, namely,\smallskip

\begin{equation}  %{Eq.36}
   |{Y}_{+^{1}\mkern-5mu/\mkern-3mu_{2}}\rangle
   =
   \frac{1}{2\sqrt{2}}
   \mkern-4mu
   \left[
      \begingroup\SmallColSep
      \begin{array}{r}
         -i\sqrt{3} \\
                    1  \\
                   -i   \\
           \sqrt{3}
      \end{array}
      \endgroup
   \right]
   \;\;\;\;  ,
\end{equation}
\smallskip

\noindent has no solution. To confirm that the ket $|{Y}_{+^{1}\mkern-5mu/\mkern-3mu_{2}}\rangle$ does not belong to $\mathrm{ran}(\hat{Y}_{+^{3}\mkern-5mu/\mkern-3mu_{2}})$, consider the statement $|{Y}_{+^{1}\mkern-5mu/\mkern-3mu_{2}}\rangle\in\mathrm{ker}(\hat{Y}_{+^{3}\mkern-5mu/\mkern-3mu_{2}})$ and present it as the system of linear equations, that is,\smallskip

\begin{equation}  %{Eq.37}
   \left[
      \begingroup
      \begin{array}{r r r}
                     7 & i\sqrt{3} &  \sqrt{3} \\
         -i\sqrt{3} &            5 &           i3 \\
           \sqrt{3} &          -i3 &            5 \\
                    -i &  \sqrt{3} & -i\sqrt{3}
      \end{array}
      \endgroup
   \right]
   \mkern-4mu
   \left[
      \begingroup\SmallColSep
      \begin{array}{r}
         x_1 \\
         x_2 \\
         x_3
      \end{array}
      \endgroup
   \right]
   =
   \frac{1}{2\sqrt{2}}
   \mkern-4mu
   \left[
      \begingroup\SmallColSep
      \begin{array}{r}
          -i\sqrt{3} \\
                      1 \\
                     -i  \\
            \sqrt{3}
      \end{array}
      \endgroup
   \right]
   \;\;\;\;  .
\end{equation}
\smallskip

\noindent Although this linear system is overdetermined (i.e., $\dim(\mathbf{X})=3<4)$, it has the solution\smallskip

\begin{equation}  %{Eq.38}
   \mathbf{X}
   =
   \frac{1}{8\sqrt{2}}
   \mkern-4mu
   \left[
      \begingroup\SmallColSep
      \begin{array}{r}
          -i\sqrt{3} \\
                     2  \\
                      i
      \end{array}
      \endgroup
   \right]
   \;\;\;\;  ,
\end{equation}
\smallskip

\noindent therefore, the statement $|{Y}_{+^{1}\mkern-5mu/\mkern-3mu_{2}}\rangle\in\mathrm{ker}(\hat{Y}_{+^{3}\mkern-5mu/\mkern-3mu_{2}})$ is true. This implies that ket $|{Y}_{+^{1}\mkern-5mu/\mkern-3mu_{2}}\rangle$ does not belong to $\mathrm{ran}(\hat{Y}_{+^{3}\mkern-5mu/\mkern-3mu_{2}})$, and so in the state $|{Y}_{+^{1}\mkern-5mu/\mkern-3mu_{2}}\rangle$ the proposition ${Y}_{+^{3}\mkern-5mu/\mkern-3mu_{2}}$ is false:\smallskip

\begin{equation}  %{Eq.39}
   {[\mkern-3.3mu[
      {Y}_{+^{3}\mkern-5mu/\mkern-3mu_{2}}
   ]\mkern-3.3mu]}_v
   =
   \mathfrak{P}_{\in}
   \!
   \left(
      |{Y}_{+^{1}\mkern-5mu/\mkern-3mu_{2}}\rangle
      ,
      \mathrm{ran}(\hat{Y}_{+^{3}\mkern-5mu/\mkern-3mu_{2}})
   \right)
   =
   0
   \;\;\;\;  .
\end{equation}
\smallskip

\noindent By contrast, in case the state of the spin $^{3}\mkern-5mu/\mkern-3mu_{2}$ system is described by the ket $|{X}_{+^{3}\mkern-5mu/\mkern-3mu_{2}}\rangle$ identified with the column vector\smallskip

\begin{equation}  %{Eq.40}
   |{X}_{+^{3}\mkern-5mu/\mkern-3mu_{2}}\rangle
   =
   \frac{1}{2\sqrt{2}}
   \mkern-4mu
   \left[
      \begingroup\SmallColSep
      \begin{array}{r}
                   1 \\
         \sqrt{3}  \\
         \sqrt{3}  \\
                  1
      \end{array}
      \endgroup
   \right]
   \;\;\;\;  ,
\end{equation}
\smallskip

\noindent neither the equations\smallskip

\begin{equation}  %{Eq.41}
   \left[
      \begingroup\SmallColSep
      \begin{array}{r}
                       1 \\
         i\sqrt{3}  \\
         -\sqrt{3}  \\
                      -i
      \end{array}
      \endgroup
   \right]
   \mkern-4mu
   x
   =
   \frac{1}{2\sqrt{2}}
   \mkern-4mu
   \left[
      \begingroup\SmallColSep
      \begin{array}{r}
                   1 \\
         \sqrt{3}  \\
         \sqrt{3}  \\
                  1
      \end{array}
      \endgroup
   \right]
   \;\;\;\;  ,
\end{equation}
\smallskip

\noindent nor the equations\smallskip

\begin{equation}  %{Eq.42}
   \left[
      \begingroup
      \begin{array}{r r r}
                     7 & i\sqrt{3} &  \sqrt{3} \\
         -i\sqrt{3} &            5 &           i3 \\
           \sqrt{3} &          -i3 &            5 \\
                    -i &  \sqrt{3} & -i\sqrt{3}
      \end{array}
      \endgroup
   \right]
   \mkern-4mu
   \left[
      \begingroup\SmallColSep
      \begin{array}{r}
         x_1 \\
         x_2 \\
         x_3
      \end{array}
      \endgroup
   \right]
   =
   \frac{1}{2\sqrt{2}}
   \mkern-4mu
   \left[
      \begingroup\SmallColSep
      \begin{array}{r}
                   1 \\
         \sqrt{3}  \\
         \sqrt{3}  \\
                  1
      \end{array}
      \endgroup
   \right]
   \;\;\;\;   
\end{equation}
\smallskip

\noindent have a solution. As a result, the ket $|{X}_{+^{3}\mkern-5mu/\mkern-3mu_{2}}\rangle$ belongs to neither $\mathrm{ran}(\hat{Y}_{+^{3}\mkern-5mu/\mkern-3mu_{2}})$ nor $\mathrm{ker}(\hat{Y}_{+^{3}\mkern-5mu/\mkern-3mu_{2}})$. In a bivalent semantics with no extra hypothesis, this implies that in the state described by $|{X}_{+^{3}\mkern-5mu/\mkern-3mu_{2}}\rangle$, the proposition ${Y}_{+^{3}\mkern-5mu/\mkern-3mu_{2}}$ is neither true nor false:\smallskip

\begin{equation}  %{Eq.43}
   {[\mkern-3.3mu[
      {Y}_{+^{3}\mkern-5mu/\mkern-3mu_{2}}
   ]\mkern-3.3mu]}_v
   =
   \mathfrak{P}_{\in}
   \!
   \left(
      |{X}_{+^{3}\mkern-5mu/\mkern-3mu_{2}}\rangle
      ,
      \mathrm{ran}(\hat{Y}_{+^{3}\mkern-5mu/\mkern-3mu_{2}})
   \right)
   =
   0/0
   \;\;\;\;  .
\end{equation}
\smallskip

\section{Lemma of partial valuation}  %{<-------------------------------------------------------------------------------------------------Section IV}

\noindent Suppose that the unit vector $|\Psi\rangle$ lies in the range of the projection operator $\hat{P}$, so that $\hat{P}|\Psi\rangle=|\Psi\rangle$. This equation corresponds to the matrix equation\smallskip

\begin{equation}  %{Eq.44}
   \mathbf{P\Psi}
   =
   \mathbf{\Psi}
   \;\;\;\;  ,
\end{equation}
\smallskip

\noindent which indicates that $\mathbf{P}^2=\mathbf{P}$. Similarly, the bra equation $\langle\Psi|=\langle\Psi|\hat{P}$ can be written as\smallskip

\begin{equation}  %{Eq.45}
   \mathbf{\Psi}^{\dagger}
   =
   \mathbf{\Psi}^{\dagger}\mathbf{P}
   \;\;\;\;  ,
\end{equation}
\smallskip

\noindent where the row vector $\mathbf{\Psi}^{\dagger}\mkern-2mu\in\mkern-2mu\mathbf{Mat_{1\times{n}}}(\mathbb{C})$, whose $i^{\text{th}}$ column has the entry $\langle\Psi|e_i\rangle$, is the adjoint matrix of $\mathbf{\Psi}$. From here, it follows that\smallskip

\begin{equation}  %{Eq.46}
   \mathbf{P\Psi}
   \mathbf{\Psi}^{\dagger}
   =
   \mathbf{\Psi\Psi}^{\dagger}\mathbf{P}
   =
   \mathbf{\Psi\Psi}^{\dagger}
   \;\;\;\;  ,
\end{equation}
\smallskip

\noindent which can be if\smallskip

\begin{equation}  %{Eq.47}
   \mathbf{P}
   =
   \mathbf{\Psi\Psi}^{\dagger}
   \;\;\;\;  .
\end{equation}
\smallskip

\noindent As $\mathbf{\Psi}$ is a $n\times1$ matrix and $\mathbf{\Psi}^{\dagger}$ is a $1\times{n}$ matrix, the above factorization means (see, for example \cite{Mirsky}) that the rank of the matrix $\mathbf{P}$ is $\mathrm{Rank}(\mathbf{P})=1$. This implies that for any $n\ge{2}$, the number of linearly independent column vectors of the matrix $\mathbf{P}$ is less than $n$, i.e.\smallskip

\begin{equation}  %{Eq.48}
   \mathrm{Rank}(\mathbf{P})
   =
   \mathrm{Span}
   \left(
      \mathbf{P}\mkern-3mu_{1}
      ,
      \dots
      ,
      \mathbf{P}\mkern-3mu_{n}
   \right)
   =
   m
   =
   1
   \;\;\;\;  .
\end{equation}
\smallskip

\noindent By the rank-nullity theorem \cite{Friedberg}, the nullity of the matrix $\mathbf{P}$ is $\mathrm{Nullity}(\mathbf{P})=n-1$; so, for any $n\ge{2}$, the number of linearly independent column vectors of the matrix $\mathbf{I}-\mathbf{P}$ is also less than $n$, namely,\smallskip

\begin{equation}  %{Eq.49}
   \mathrm{Nullity}(\mathbf{P})
   =
   \mathrm{Span}
   \left(
      \mathbf{I}_{1}
      -
      \mathbf{P}\mkern-3mu_{1}
      ,
      \dots
      ,
      \mathbf{I}_{n}
      -
      \mathbf{P}\mkern-3mu_{n}
   \right)
   =
   k
   =
   n
   -
   1
   \;\;\;\;  .
\end{equation}
\smallskip

\noindent Consequently, both linear systems, $\mathbf{RX}=\mathbf{\Psi}$ and $\mathbf{KX}=\mathbf{\Psi}$, are overdetermined.\\

\noindent Let’s prove the lemma of partial valuation:\\

\noindent \textbf{Lemma:} \emph{With no extra hypothesis, algebraic properties of the separable Hilbert space $\mathcal{H}$ of finite dimension $n\ge2$ cannot allow in valuations $v$ which are total functions from $\mathbb{P}$ to $\mathbb{B}_2$}.\\

\noindent Suppose, the opposite is true, namely, valuations $v$ that are total functions from $\mathbb{P}$ to $\mathbb{B}_2$ are admitted. Then, for every nonzero vector $|\Psi\rangle$ in $\mathcal{H}$ and an arbitrary projection operator $\hat{P}$ such that $\hat{P}\neq\hat{0}$ and $\hat{P}\neq\hat{1}$, it holds true that $|\Psi\rangle$ belongs to either $\mathrm{ran}(\hat{P})$ or $\mathrm{ker}(\hat{P})$. Then again, this can happen only if at least one solution to either system of equations, $\mathbf{RX}=\mathbf{\Psi}$ or $\mathbf{KX}=\mathbf{\Psi}$, is guaranteed regardless of either $\mathbf{R}$ and $\mathbf{\Psi}$, or $\mathbf{K}$ and $\mathbf{\Psi}$, that is, in case either $m$ or $k$ is equal to $n$.\\

\noindent But, because both systems are overdetermined, this cannot be guaranteed: $\mathbf{RX}=\mathbf{\Psi}$ and $\mathbf{KX}=\mathbf{\Psi}$ will necessarily be unsolvable for some choice of values for the left-hand and right-hand sides of their equations, i.e., for some couples $(|\Psi\rangle,\mathrm{ran}(\hat{P}))$.\\

\noindent Stipulating that $U_{\mathbf{R}}$ is the solution set for the linear system $\mathbf{RX}=\mathbf{\Psi}$\smallskip

\begin{equation}  %{Eq.50}
   U_{\mathbf{R}}
   =
   \left\{
      \mathbf{X}
      \mkern-2mu
      \in
      \mkern-2mu
      \mathbf{Mat_{m\times{1}}}(\mathbb{C})
      \mkern-3mu
      :
      \mkern7.5mu
      \mathbf{RX}
      =
      \mathbf{\Psi}
   \right\}
   \;\;\;\;  ,
\end{equation}
\smallskip

\noindent and $U_{\mathbf{K}}$ is the solution set for the linear system $\mathbf{KX}=\mathbf{\Psi}$\smallskip

\begin{equation}  %{Eq.51}
   U_{\mathbf{K}}
   =
   \left\{
      \mathbf{X}
      \mkern-2mu
      \in
      \mkern-2mu
      \mathbf{Mat_{k\times{1}}}(\mathbb{C})
      \mkern-3mu
      :
      \mkern7.5mu
      \mathbf{KX}
      =
      \mathbf{\Psi}
   \right\}
   \;\;\;\;  ,
\end{equation}
\smallskip

\noindent this inference can be written by the series of the sentences:\smallskip

\begin{equation}  %{Eq.52}
   \begingroup\SmallColSep
   \begin{array}{r l}
      |\Psi\rangle
      \in
      \mathrm{ran}
         (\hat{P})
      &
      \iff
      U_{\mathbf{R}}\neq\emptyset
      \\
      |\Psi\rangle
      \in
      \mathrm{ker}
         (\hat{P})
      &
      \iff
      U_{\mathbf{K}}\neq\emptyset
      \\
      |\Psi\rangle
      \in
      \mathrm{ran}
         (\hat{P})
      \downarrow
      |\Psi\rangle
      \in
      \mathrm{ker}
         (\hat{P})
      &
      \iff
      U_{\mathbf{R}}\neq\emptyset
      \downarrow
      U_{\mathbf{K}}\neq\emptyset
   \end{array}
   \endgroup   
   \;\;\;\;  ,
\end{equation}
\smallskip

\noindent where the symbol $\iff$ stands for \emph{the logical biconditional} (which is true when its antecedent and consequent are either true or false at the same time).\\

\noindent According to the rank-nullity theorem, $U_{\mathbf{R}}\neq\emptyset \downarrow U_{\mathbf{K}}\neq\emptyset$ may be true (meaning that both solution sets $U_{\mathbf{R}}$ and $U_{\mathbf{K}}$ may be empty) for any $n\ge2$. Thus, without additional hypotheses, the valuations $v$, which are total functions mapping experimental quantum propositions to elements of $\mathbb{B}_2$, cannot be admitted for any separable Hilbert space $\mathcal{H}$ of finite dimension $n\ge2$.\\

\section{Closing remarks}  %{<-------------------------------------------------------------------------------------------------Section V}

\noindent Consider the statement (which can be called experimental quantum proposition) “A physical quantity of a quantum system has a certain value”. \emph{Of what is this statement a primary bearer?}\\

\noindent The answer, that is ordinarily being given in the literature, is that this sentence is \emph{a primary bearer of probability values}. This answer suggests that truth values of experimental quantum propositions can be regarded as degenerate probabilities. Consequently, rather than truth preservation, the logic of experimental quantum propositions is about probability preservation.\\

\noindent Given that the mathematical representative of any experimental quantum proposition is a closed linear subspace of a Hilbert space $\mathcal{H}$, this implies that quantum mechanics can be reduced to the theory of probability measures on closed linear subspaces of $\mathcal{H}$. According to Gleason’s theorem, if the Hilbert space has a finite dimension $n\ge3$, this theory does not admit probability measures having only the values 0 and 1. The above result can be interpreted as evidence ruling out the possibility of hidden variables in quantum mechanics.\\

\noindent In contrast, in the present paper, another answer is offered asserting that an experimental quantum proposition is a primary bearer of truth values. Upon that, the paper demonstrates that without additional suppositions, algebraic properties of separable Hilbert spaces $\mathcal{H}$ of finite dimension $n\ge2$ do not allow in truth assignments $v$ which are dispersion-free, i.e., which have being total functions from the set of atomic (i.e., elementary) experimental quantum propositions to the set of two objects, true and false (or 1 and 0, correspondingly).\\

\noindent As long as a probability is interpreted as a measure of the (un)certainty in the assignment of truth values to an atomic proposition, the fact that the valuations $v$ cannot be dispersion-free indicates that probabilities cannot be only $\{0,1\}$-valued. In this way, one can say that gaps in truth assignments entail the emergence of probabilistic semantics for the logic of experimental quantum propositions.\\

\noindent It follows – independently of Gleason’s theorem – that non-contextual hidden variables identified as globally defined $\{0,1\}$-valued observables (resembling experimental propositions) must be excluded from the interpretation of quantum mechanics that is based on the Hilbert space formalism.\\

\bibliographystyle{References}

\end{document}